\documentclass[11pt]{article}
\pdfoutput=1
\usepackage{amsmath}
\usepackage{amsfonts}
\usepackage{hyperref}
\usepackage{caption}
\usepackage{fullpage}
\usepackage{subcaption}
\usepackage{graphicx}
\usepackage{authblk}
\usepackage[T1]{fontenc}
\usepackage[utf8]{inputenc}

\date{}
\begin{document}

\title{An integrative statistical model for inferring strain admixture within clinical \emph{Plasmodium falciparum} isolates}

\author[1,*]{John D O'Brien}
\author[2]{Zamin Iqbal}
\author[2,3]{Lucas Amenga-Etego}
\affil[1]{Department of Mathematics, Bowdoin College, Brunswick, Maine, USA}
\affil[2]{Wellcome Trust Centre for Human Genetics, University of Oxford, Oxford, UK}
\affil[3]{Navrongo Health Research Centre, Navrongo, Upper East Region, Ghana}
\affil[*]{To whom correspondence should be addressed}

\maketitle

\section*{ABSTRACT}
Since the arrival of genetic typing methods in the late 1960's, researchers have puzzled at the clinical consequence of observed strain mixtures within clinical isolates of \emph{Plasmodium falciparum}. We present a new statistical model that infers the number of strains present and the amount of admixture with the local population (panmixia) using whole-genome sequence data. The model provides a rigorous statistical approach to inferring these quantities as well as the proportions of the strains within each sample. Applied to 168 samples of whole-genome sequence data from northern Ghana, the model provides significantly improvement fit over models implementing simpler approaches to mixture for a large majority (129/168) of samples. We discuss the possible uses of this model as a window into within-host selection for clinical and epidemiological studies and outline possible means for experimental validation.

\section*{INTRODUCTION}

The protozoan parasite \emph{Plasmodium falciparum} (Pf) is the cause of the vast majority of fatal malaria cases, killing at least half a million people a year \cite{Hay2009,Snow2005,World2008}. The parasite's ability to develop resistance to drugs and the rapid spread of that resistance across geographically-separated populations presents a constant threat to international control efforts \cite{Wootton2002,Mita2009,Payne1987}. While research has elucidated many genetic factors in resistance, much of genetic epidemiology of the parasite - including the effective recombination rate and the rate of gene flow across populations - is still unclear \cite{Sidhu2002,Mita2009,Roper2004}.  

Since the late 1960's, researchers focused on the structure of Pf clinical infections have struggled to understand the implications of multiplicity of infection (MOI), where multiple strains appear to be present within a single patient's bloodstream \cite{Wilson1969,Mcgregor1972,Jamjoom1983,Conway1991,Auburn2012}. While MOI-focused studies implicate increased or decreased levels of MOI with a range of conditions, including clinical severity \cite{Muller2001}, age-specific severity \cite{Henning2004,Smith1999a,Farnert1999,Stirnadel1999}, parasitemia levels during pregnancy \cite{Beck2001}, and other effects \cite{Beck1999,Smith1999b,Paganotti2004,Mayengue2009}, there is no broad consensus about MOI's role, if any, in controling the course of an infection. Still, a wide variety of studies and genetic assays -- most typically by typing of the \emph{mdr} gene -- show MOI as a reliable feature of clinical Pf isolates \cite{Auburn2012}.  

The appearance of whole-genome sequencing (WGS) technologies applied to Pf extracted directly from infected patients' bloodstreams provides an unprecedented window into the structure of genetic mixture within samples \cite{Manske2012,Auburn2011}. Initial work on understanding the structure of this mixture data shifted focus from estimating MOI to analysis based on inbreeding coefficients \cite{Auburn2012,O'Brien2014,Weir1984}. These metrics, a special type of $F$-statistic, provide an estimate of the departure of within-sample allele frequencies from those expected under a Hardy-Weinberg-type equilibrium with the nearby population. In this perspective, each patient's bloodstream is taken to be a subpopulation exhibiting a degree of admixture of all of the strains from the local environment, ranging from a perfectly random sampling of all nearby strains to the repeated sampling of just single strain. 

The initial study applying WGS to clinical Pf isolates collected from eight countries on three continents shows that the parasite exhibits significant population structure at continental scales, with the amount of subpopulation structure varying significantly among regions \cite{Manske2012}. Employing novel F-statistics to measure the inbreeding coefficient, this work also argued that the degree of mixture varies significantly across populations, with highly mixed samples occurring relatively frequently in west Africa but only occasionally in Papua New Guinea. Importantly, the authors suggested an association between increased levels of observed mixture with increases in transmission intensity in the local environment. Transmission intensity, the rate at which individuals are infected with Pf, likely determines some part the frequency of out-crossing within parasite populations and so may be critical to understanding gene flow and strategies for resistance control \cite{Guerra2008}. 

In this paper, we present a new statistically rigorous model that synthesizes these two distinct and previously disparate approaches to analyzing \emph{P. falciparum} clinical mixtures: assessing the number of distinct genetic types within a sample (the MOI approach) and measuring the degree of panmixia with respect to the local population (the inbreeding coefficient approach). The model centers around how these two sub-models contribute to generate the observed within-sample non-reference allele frequency as it relates to the population-level non-reference allele frequency for single nucleotide polymorphisms (SNPs). For clarity, we will deprecate the use of \emph{non-reference} in front of the term allele frequency, since they are all calibrated in this fashion. We will use the acronyms WSAF to denote the within-sample allele frequency and PLAF to denote population-level allele frequency to avoid confusion about the particularly frequency being indicated.

\begin{figure}[!ht]
\begin{center}
\includegraphics[scale=0.45]{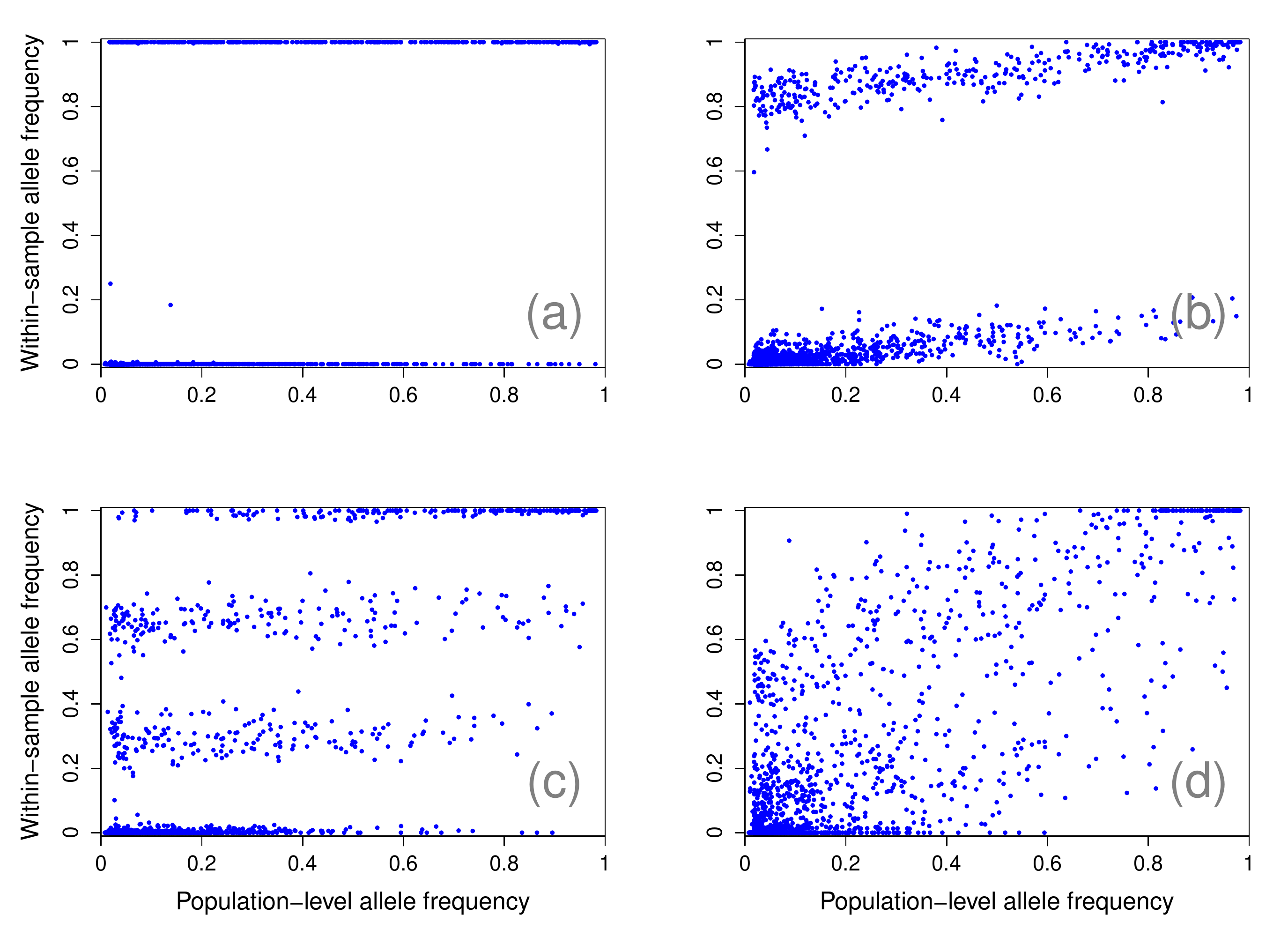}
\end{center}
\caption{Four representative samples with WSAF for each SNP plotted against the PLAF, showing an absence of mixture (a), a partially panmixed sample (b), a simple mixture (c), and a complex mixture (d).}
\label{example}
\end{figure}

The essential structure of the model is to explain observed `bands' that emerge when examining the WSAF for SNPs as a function of their PLAF (Figure \ref{example}). The model posits that the number of these bands results as a direct consequence of the number of distinct strains present within a sample and that the degree of admixture with the local population determines bands' slopes. To distinguish from the inbreeding coefficient, we refer to the degree of admixture as the panmixia coefficient. The collection of bands are then modeled jointly as a finite mixture. 

\begin{figure}[!ht]
\begin{center}
\includegraphics[scale=0.45]{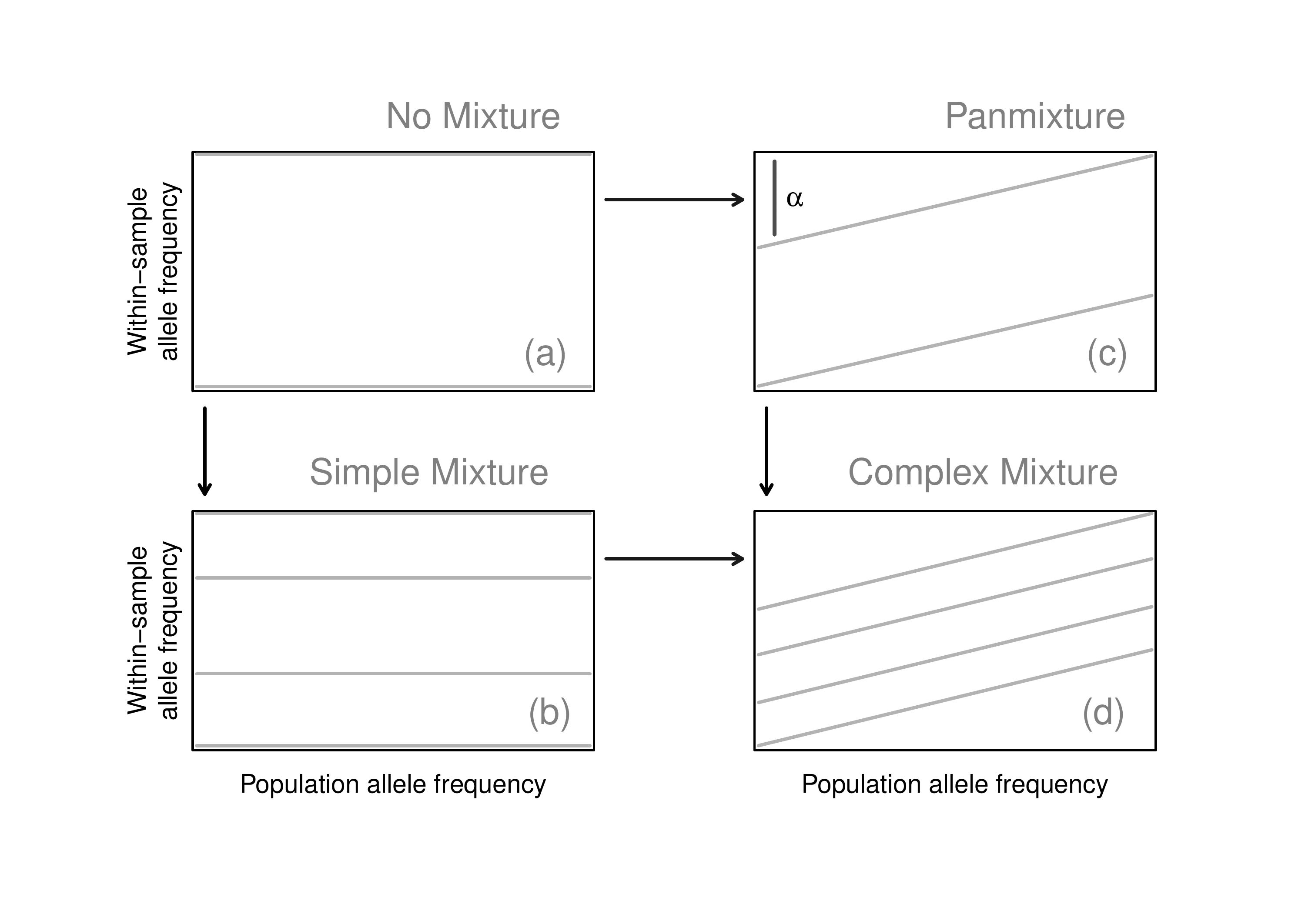}
\end{center}
\caption{The essential structure of the model comprises four distinct states, relating the WSAF to the PLAF: no mixture (upper left); simple mixture (lower left); panmixture (upper right); and complex mixture (lower right).}
\label{diagram}
\end{figure}

Figure \ref{diagram} lays out the important components of the model. In the simplest case a sample is composed of a single, unmixed strain, and all SNPs exhibit a WSAF of zero or one (see Figure \ref{diagram}(a)), depending on whether they agree with the reference.  Consequently the WSAF is independent of PLAF, leading to two flat bands at these values. We call these samples unmixed. In the case where finite number of strains mixed within a sample, then for each variant position some number of those strains will possess a reference allele and some will not.  Which strains carry non-reference alleles and those strains' proportion in the sample mixture then determine the WSAF for each SNP. Observed across many SNPs, this leads to the apparent bands of constant WSAF across the PLAF. It follows that for $K$ component strains there are $2^K$ possible combinations of biallelic states, leading to that number of apparent WSAF bands. 

The complementary banding structure of panmixia arises when a fraction of the Pf organisms present within the blood are randomly sampled from the local population. In its simplest formulation, the panmixture model represents the admixture of two distinct Pf populations: a single strain, representing $1 - \alpha$ of the within-sample genomes, and a random sample of strains from the local population, representing $\alpha$ of the remaining genomes. In the case of perfect panmixia ($\alpha=1$), a sample would be comprised of organisms evenly sampled from the ambient population and the plot of WSAF against PLAF would become a single line at $y=x$. In this data set, we do not observe any sample close perfect panmixia but observe several instances of apparent partial panmixia with a single dominant strain (Figure \ref{example}(c) and \ref{diagram}(c)). The $\alpha$ tilt in the WSAF arises from the fact that for this proportion of organisms the probability of sampling non-reference allelele is proportional to the PLAF, absent any other population structure. These samples, with a single strain and a degree of panmixia, we call panmixed. In more complex cases, where there is more than one dominant strain, the total number of bands is still determined by the number of these strains. However, now panmixia tilts each of these bands equally leading to complex mixtures (Figures  \ref{example}(d) and \ref{diagram}(d)).
The paper proceeds as follows. We first detail the structure of the WGS data, introduce some notation, and the essential mathematical structure of the model. We then present an extensive simulation study on the performance of the model and then an examination of its application to field isolates collected from northern Ghana. We conclude by discussing the strengths and weakness of the model, some possible improvements, and what consequences this analysis may have for understanding the etiology of clinical malaria.

\section*{DATA, NOTATION, AND MODEL}

\subsection*{Data}
The WGS data come from Illumina HiSeq sequencing applied to \emph{P. falciparum} extracted from $235$ clinical blood samples collected from infected patients from the Kassena-Nankana district (KND) region of Upper East region of northern Ghana. Collection occurred over approximately $2$ years, from June 2009 to June 2011. The full sequencing protocol and collection regime are described in \cite{Manske2012}. After quality control measures, sequencing was performed on $235$ samples, and, following a documented protocol using comparison against world-wide variation, $198,181$ single-nucleotide polymorphisms (SNPs) were called within each sample \cite{Manske2012}. Each call provides the number of reference and non-reference read counts observed at each variant position within the genome, ascertained against the the $3\mbox{D}7$ reference \cite{Gardner2002}. For this project, we additionally filtered these data. First, multiallelic positions were reclassed as biallelic. We then excluded positions that exhibited no variation within the KND samples, any level of missingness (no read counts observed), or minor allele frequency less than $0.01$. To remove low quality samples, we removed thoses possesed more than $4,000$ SNPs called with fewer than $20$ read counts, following an inflection point observed in Supplementary Figure S1(a). These cleaning measures left $2,429$ SNPs in $168$ samples. More than $95\%$ of remaining samples' sequencing was completed without PCR amplification. We observe little apparent population structure among the samples, evidenced either by principal components analysis or a neighbor-joining tree of the pairwise difference among samples (Supplementary Figures S2). The data preparation scripts are available with the source code for the model, \href{https://github.com/jacobian1980/pfmix/}{https://github.com/jacobian1980/pfmix/}.  

\subsection*{Notation}

Following the data preparation and cleaning, our analysis begins with a set of $N$ clinical samples, each composed of $M$ SNPs. At each SNP $j$ within each clinical sample $i$, we observe $r_{ij}$ reads that agree with the reference genome and $n_{ij}$ reads that do not agree with the reference. For a sample $j$. we write the complete data across all SNPs as $\mathcal{D}_i = [(r_{i1},n_{i1}),\cdots,(r_{iM},n_{iM})]$. For each SNP $j$, we associate a PLAF $p_j$. The collection of all $p_j$ we refer to as $\mathcal{P}$.

Conditional upon the number of strains $K$, there are $2^K$ bands, indexed by $r=1,\cdots,2^K$.  The full collection of bands we call $\mathcal{Q}$, with $q_{ijr}$ indicating the WSAF for band $r$ at SNP $j$ in sample $i$. The probability of a SNP lying within the distinct bands across the PLAF is specified by a mixture component $\lambda_r$, which is a function of the PLAF, and so is often written $\lambda_r(p_j)$. The degree of panmixia in a sample is given by $\alpha$, a value between zero and one. A complete list of the model parameters is given in Table \ref{notation}.

\subsection*{Model}

Statistically, the model takes the form of a finite mixture model, with the mixture components associated with individual bands \cite{Redner1984,Mclachlan2004}. We take a Bayesian approach to inference and so lay out the model by giving an overall rationale for the decomposition of the posterior distribution and then justifying the appropriate choice of probability distributions for each of the terms \cite{Gelman2013}. 
\subsubsection*{Decomposition}
We assume that samples are independent of each other and that the SNP data for each sample depends solely on $K$, the WSAF $\mathcal{Q}$, the PLAF $\mathcal{P}$, and a shape parameter $\nu$. As samples are independent, we will deprecate sample-specific subscripts for the model parameters. Considering the data for a single sample, $\mathcal{D}_i$, the posterior distribution can then be written as: 
\begin{eqnarray}
\mathbb{P}(\mathcal{Q},\mathcal{P},  \mathcal{W}, \alpha, \nu, K | \mathcal{D}_i)  & \propto & \mathbb{P}(\mathcal{D}_i | \mathcal{Q}, \mathcal{P}, \mathcal{W}, \alpha, \nu, K ) \cdot \mathbb{P}(\mathcal{Q},\mathcal{P}, \mathcal{W}, \alpha, \nu, K) \\
& = & \mathbb{P}(\mathcal{D}_i | \mathcal{Q}, \mathcal{P}, \nu, K) \cdot  \mathbb{P}(\mathcal{Q}, \mathcal{P}, \nu, K, \mathcal{W}, \alpha) \mbox{.} 
\label{init_decomp}
\end{eqnarray}

We also assume that the WSAF, $\mathcal{Q}$, depends only on the PLAF, $\mathcal{P}$, the panmixia coefficient $\alpha$, the number of strains $K$, and their proportions within the sample, $\mathcal{W}$, allowing the right-hand side of Equation \ref{init_decomp} to be further decomposed, by noting that
\begin{eqnarray}
\mathbb{P}(\mathcal{Q}, \mathcal{P}, \nu, K, \mathcal{W}, \alpha)  & = &  \mathbb{P}(\mathcal{Q}|\mathcal{P}, \nu, K, \mathcal{W}, \alpha) \cdot \mathbb{P}(\mathcal{P}, \nu, K, \mathcal{W}, \alpha) 
\label{second_decomp} \mbox{.}
\end{eqnarray}
While $\mathcal{W}$ clearly depends on the number of strains, $K$, the remaining parameters we take to be independent of this value and of each other. This means that the last right-hand side term in Equation \ref{second_decomp} becomes:
\begin{eqnarray}
\mathbb{P}(\mathcal{P}, \nu, K, \mathcal{W}, \alpha)  & = &  \mathbb{P}(\mathcal{P}) \cdot \mathbb{P}(\nu) \cdot \mathbb{P}(\mathcal{W}|K) \cdot \mathbb{P}(K) \cdot \mathbb{P}(\alpha)  \mbox{.}
\label{third_decomp}
\end{eqnarray}
Substituting Equations \ref{second_decomp} and \ref{third_decomp} into Equation \ref{init_decomp}, yields the final decomposition:
\begin{eqnarray}
\mathbb{P}(\mathcal{Q},\mathcal{P},  \mathcal{W}, \alpha, \nu, K | \mathcal{D}_i)  & \propto & \mathbb{P}(\mathcal{D}_i | \mathcal{Q}, \mathcal{P}, \nu, K) \cdot \mathbb{P}(\mathcal{Q}|\mathcal{P}, \nu, K, \mathcal{W}, \alpha) \cdot \nonumber \\
& & \hspace{1.5cm}\mathbb{P}(\mathcal{P}) \cdot \mathbb{P}(\nu)  \cdot \mathbb{P}(\mathcal{W}|K) \cdot \mathbb{P}(K) \cdot \mathbb{P}(\alpha)     \mbox{.} 
\end{eqnarray}
We now specify each of the terms on the right-hand side above as probability distributions.  
\subsubsection*{Likelihood :  $\mathbb{P}(\mathcal{D}_i | \mathcal{Q}, \mathcal{P}, \nu, K)$}
Within band $r$, the WSAF at SNP $j$ in sample $i$ is $q_{ijr}$.  Supposing that read counts at $j$ are identically and independently distributed with probability $q_{ijr}$, we model the probability of the data $(r_{ij},n_{ij})$ as a Beta-binomial distribution, allowing us to model greater dispersion than expected under a pure binomial. We parameterize this distribution in terms of $q_{ijr}$ and $\nu$ rather than the more commonly used shape and scale parameters, $\alpha$ and $\beta$.  The relationship between the two parameterization is $q_{ijr} \cdot \nu = \alpha$ and $(1-q_{ijr}) \cdot \nu = \beta$. We use this parameterization as it allows us to write the model in terms of the mean allele frequency that defines each band. The additional  $\nu$ is a shape parameter that serves as a proxy for the variance. These parameters give a likelihood expression:
\begin{eqnarray}
\mathbb{P}(n_{ij}, r_{ij}|r, q_{ijr}, \nu) &=& {n_{ij} + r_{ij} \choose n_{ij} } \cdot \frac{\mbox{B}(n_{ij} + q_{ijr} \cdot \nu,r_{ij} + (1-q_{ijr})\cdot \nu)}{\mbox{B}( q_{ijr}\cdot \nu, (1-q_{ijr})\cdot \nu)} \mbox{,}
\label{likelihood}
\end{eqnarray}
where $\mbox{B}$ is the beta function. 

As any SNP could lie within any band, we employ a novel version of the finite mixture model to capture this segregation. Fixing the number of strains to $K$, there are then $2^K$ ways that the strains can be assorted into non-reference and reference allele states at any given position $j$. A given band $r$ arises from $C_r$ strains exhibiting the non-reference allele and $2^K - C_r$ strains exhibiting the reference allele.  Supposing no population structure among the strains, the probability that a given SNP will be in that band is simply the probability of drawing $C_r$ non-reference alleles and $2^K-C_r$ reference alleles, conditional upon $p_j$:
\begin{eqnarray*}
\mathbb{P}(\mbox{SNP }j \in \mbox{band } r|p_j) &=& p_j^{C_r}  \cdot (1- p_j)^{2^K-C_r} \\ 
& = & \lambda_r(p_j) \mbox{.}
\label{lambda}
\end{eqnarray*}
Consequently, the density of the mixture coefficients for each band varies across the PLAF but such that they sum to $1$ across all bands at any position $j$:
\begin{eqnarray*}
\mathbb{P}(\mathcal{D}_{ij} | \mathcal{Q}, \mathcal{P}, \nu, K) & = &  \sum_{r=1}^{2^K} \mathbb{P}(\mbox{SNP }j \in \mbox{band } r|p_j) \cdot \mathbb{P}(n_{ij}, r_{ij}|r, q_{ijr}, \nu)  \\
& = & \sum_{r=1}^{2^K} \lambda_r(p_j) \cdot \mathbb{P}(n_{ij}, r_{ij}|r, q_{ijr}, \nu)  \mbox{.}
\end{eqnarray*}
  Following from the assumption of no population structure, SNPs will assort into bands independently. This leads to a product form for the likelihood of sample's data, $\mathcal{D}_i$:

\begin{eqnarray}
\mathbb{P}(\mathcal{D}_i | \mathcal{Q}, \mathcal{P}, \nu, K) & = & \prod_{j=1}^M \bigg[  \sum_{r=1}^{2^K} \lambda_r(p_j) \cdot \mathbb{P}(n_{ij}, r_{ij}|r, q_{ijr}, \nu) \bigg] \mbox{.}
\label{likelihood_2}
\end{eqnarray}

\subsubsection*{Band structure: $\mathbb{P}(\mathcal{Q}|\mathcal{P}, \nu, K, \mathcal{W}, \alpha) $}

The full mixture model contains two distinct subcomponents that we call the simple mixture model and the panmixture model, respectively. Both models generalize the unmixed case, though naturally in different ways. We first describe the unmixed model and then layout the two extensions before showing how these can be combined to create the full model. In practice, we only fit data using the full model and allow it to indicate the number of strains, their proportions, and the degree of panmixia.  We do not know the number of strains \emph{a priori} so we employ metrics applied to the posterior distribution inferred with different values of $K$ to determine it. However, for the purpose of detailing the model, we assume that $K$ is known. \\

\noindent \textbf{Unmixed model} - In an unmixed sample only one strain is present and there is no panmixia, and so $K=1$ and $\alpha=0$. Consequently, we expect all SNPs to exhibit WSAF either zero or one (Figure \ref{diagram}(a)) depending on whether they agree with the reference or not.  There are then two bands, $r=1,2$ and $q_{ij1} = 0$ and $q_{ij2} = 1$.  \\

\noindent \textbf{Simple mixture model} - The simple mixture model assumes that a finite number $K$ of distinct strains, $s_1, \cdots, s_K$, are combined together in the sample with proportions, $\mathcal{W} = [w_1,\cdots, w_K]$ but that $\alpha=0$.  Naturally, $\sum_{k} w_k = 1$ and for each SNP $j$, the probability of being within band $r$ is given by $\lambda_r(p_j)$, as above.  Band $r$ is defined by a vector $v_r = [\mathbf{1}_{\{s_1 \in r\}}, \cdots, \mathbf{1}_{\{s_K \in r\}}]$, where $\mathbf{1}_{\{s_k \in r\}}$ is an indicator function of whether strain $k$ exhibits a non-reference allele within the sample.  The WSAF of $q_{ijr}$ is then given by the sum of all of proportions of strains that exhibit a non-reference allele:  
\begin{eqnarray}
q_{ijr} &=& \sum_{k=1}^K w_k \cdot \mathbf{1}_{\{s_k \in r\}}  \mbox{.}
\label{simple_model}
\end{eqnarray}
Taken across all $r$ bands, this leads to $2^K$ bands with zero slope and corresponding proportions $(0,w_1,\cdots,w_K,w_1+w_2,w_1+w_3,\cdots,1)$.

\noindent \textbf{Panmixture model} - As mentioned above, in its simplest case, the panmixture model represents the admixture of two distinct Pf populations: a single strain, representing $1-\alpha$ of the within-sample orgnaisms, and a random sample of strains from the local population, for the remaining $\alpha$ organisms. When $\alpha = 0$ the model reduces to the unmixed case. We will refer the single strain as the dominant strain, although, conditional upon $\alpha$, it may represent only a small proportion of the sample's population. For each position $j$, there are still only two bands: the higher one corresponding to the non-reference allele being present in the dominant strain, and the lower one corresponding to its absence. However, the WSAF for these bands varies according to $p_j$ with slope $\alpha$. To resolve $q_{ijr}$, first consider the upper band, $r=2$.  At any position $j$, $1-\alpha$ of the reads come from the dominant strain.  The remaining reads, each sampled randomly from the local population, each have probability $p_j$ of being non-reference. This leads to $q_{ij2} =  (1-\alpha) + \alpha \cdot p_j$. For the lower band, the dominant strain contributes no non-reference reads so $q_{ij1} =   \alpha \cdot p_j$.\\

\noindent \textbf{Complex mixture model} - The complex model synthesizes the simple mixture and panmixture models.  In this case, at position $j$, $\alpha$ of the reads are sampled randomly from the across the local population, contributing a fraction of $\alpha \cdot p_j$ non-reference alleles. The state of the remaining reads are determined by $\mathcal{W}$ as in Equation $\ref{simple_model}$. For band $r$ at position $j$, the WSAF is then given by 
\begin{eqnarray}
q_{ijr} & = &  (1-\alpha) \cdot \bigg(  \sum_{k=1}^K w_k \cdot \mathbf{1}_{\{s_k \in r\}}  \bigg)+ \alpha \cdot p_j \mbox{.}
\label{full_model}
\end{eqnarray}
There are then $2^K$ bands with proportions $(0,w_1,\cdots,w_K,w_1+w_2,w_1+w_3,\cdots,1)$ and slope $\alpha$.

\subsubsection*{Priors}
For the remaining four probability distributions we place the following vague prior distributions:
\begin{eqnarray*}
\mathcal{W}|K & \sim & \mbox{\small{DIRICHLET}}(\mathbf{1}_K) \\
\alpha & \sim & \mbox{\small{UNIFORM}}(0,1) \\
\nu & \sim & \mbox{\small{EXPONENTIAL}}(5)\\
K & \sim & \mbox{zero-truncated } \mbox{\small{POISSON}}(2) \mbox{,}
\end{eqnarray*}
where $\mathbf{1}_K$ is a vector of $K$ ones. 

\subsection*{Inference}
We infer the model parameters using a standard Bayesian Markov chain Monte Carlo (MCMC) approach \cite{Gilks2005,Geyer1992} with one exception: we first calculate maximum-likelihood estimates (MLE) for $\mathcal{P}$ across all samples and then treat these as fixed when inferring the remaining parameters \cite{Scholz1985}. This choice is motivated by statistical expedience and computational speed. Except for $\mathcal{P}$, the parameters of the model are independent across samples and so this approximation enables the algorithm to infer parameters in parallel rather than jointly. This avoids the difficulties of performing inference on the number of strains within samples simultaneously, which would require an involved trans-dimensional MCMC scheme (such as reversible jump MCMC, \cite{Green1995}) acting jointly across all samples. Running in parallel also increases the computational speed of the implementation by at least an order of magnitude. Since the sample collection is large enough that $\mathcal{P}$ is nearly independent of any given sample, we do not expect this approximation to significantly bias inference.

For each SNP $j$, the MLE derives from treating the non- and reference reads within a sample as coming from a binomial distribution with parameter $p_j$. This leads to:  
\begin{eqnarray*}
\hat{p}_j & = & \displaystyle \sum_{i}^N n_{ij} \bigg/ \displaystyle \sum_{i}^N (n_{ij} + r_{ij})\mbox{.}
\end{eqnarray*}
To infer $K$ for each sample, we employ a Bayesian Information Criterion (BIC) \cite{Posada2004,Chen1998} and harmonic mean estimator to the Bayes Factor (hmeBF) \cite{Lavine1999,Kass1995} as metrics for model selection. To find the maximum likelihood value for use with the BIC, we initially implemented a separate estimation algorithm but found no significant difference with using the highest value observed from the posterior samples. In simulations, we observe that the BIC and hmeBF provide similar guidance, with the BIC frequently indicating a smaller $K$. For the simulation study and empirical data example, we provide only the BIC result.

Conditional on $\mathcal{P}$ and $K$, we implement a Metropolis-Hastings algorithm to draw samples from the posterior distribution \cite{Gilks2005}. For each of the three parameters, $\alpha$, $\mathcal{W}$, and $\nu$, we propose new values directly from the prior distribution, leading to Metropolis-Hastings ratios almost solely dependent on the ratio between the likelihood and priors for the proposed state to those for the current. The inference scheme is implemented in set of scripts for the R computing language, and can be found under the Academic Free License at \href{https://github.com/jacobian1980/pfmix/}{https://github.com/jacobian1980/pfmix/s}. For a single sample, a sufficiently long MCMC run takes approximately 20 minutes on a single high-performance computing core.

\section*{RESULTS}
\subsection*{Simulations under the model}
To demonstrate the efficacy of our implementation, we present a simulation study examining the algorithm's performance. We consider two distinct aspects of the inference separately: how well the model infers the number of strains, and, conditional upon that number, how well it infers the model's other parameters.  We simulate data from the model in the following way. Conditional upon $M$,$\alpha$, $K$ and the sum of the read counts, $C$, we draw a vector of probabilities, $\mathcal{W}$, from a uniform Dirichlet distribution. We combine the values of $\mathcal{W}$ in all possible permutations to create the $2^K$ bands and assign the PLAF for the SNPs evenly from $1/M$ to $1$, so that the $j^{\mbox{\tiny{th}}}$ SNP has PLAF $\frac{j}{M}$. For each SNP, we first probabilistically select the band it occupies according to according to Equation \ref{lambda}. Conditional upon selecting $r$, we then simulate read counts according to the likelihood (Equation \ref{likelihood}) with $q_{ijr}$ according to  Equation \ref{full_model}. For all simulations, we set $\nu=10$. We run the simulation across the range of values for $M$,$\alpha$, $K$ and $C$. For each parameter set, we create $10$ independent realizations. 

\subsubsection*{Number of components}
\begin{figure}[!ht]
\begin{center}
\begin{tabular}{cc}
\includegraphics[scale=0.3]{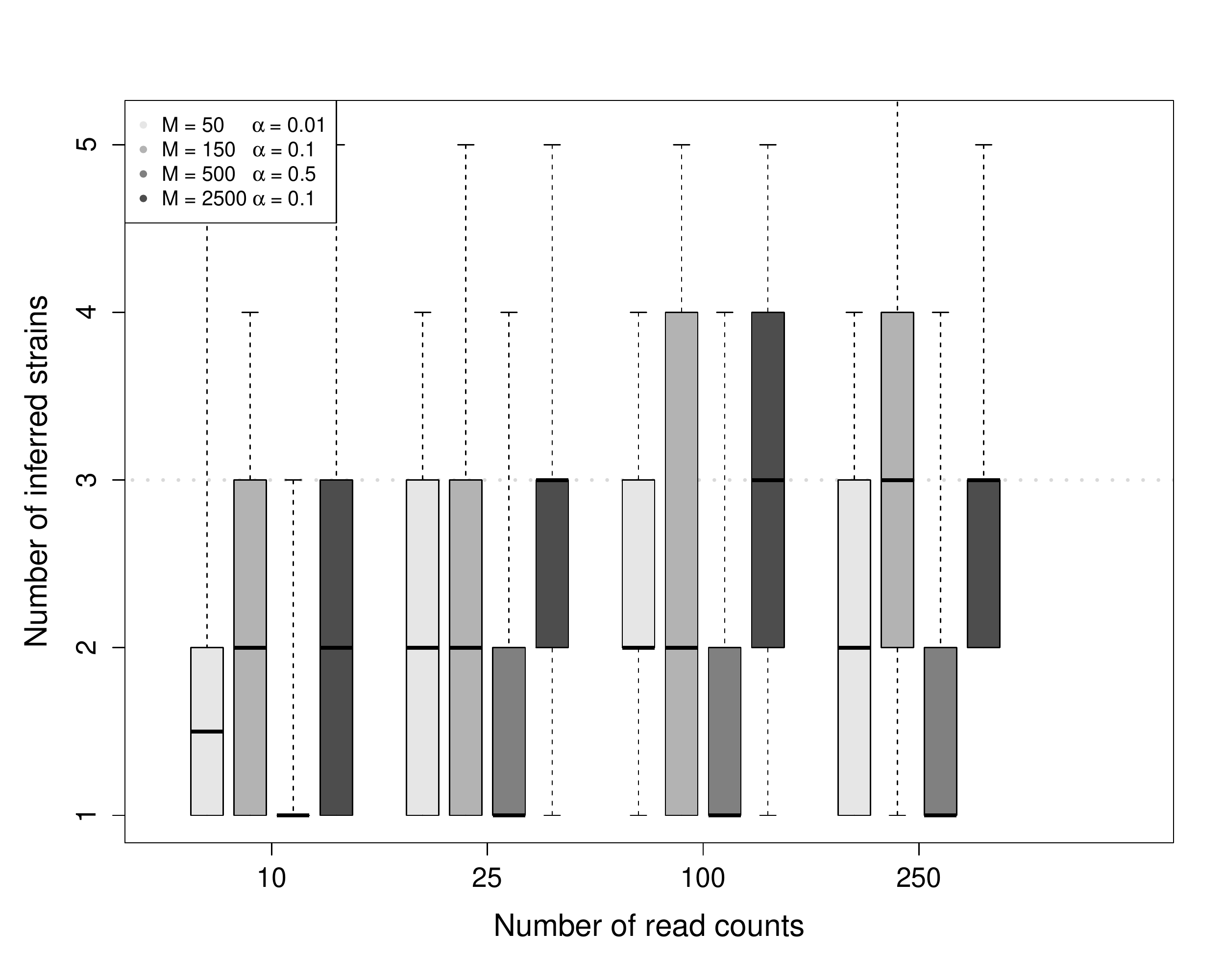} &
\includegraphics[scale=0.3]{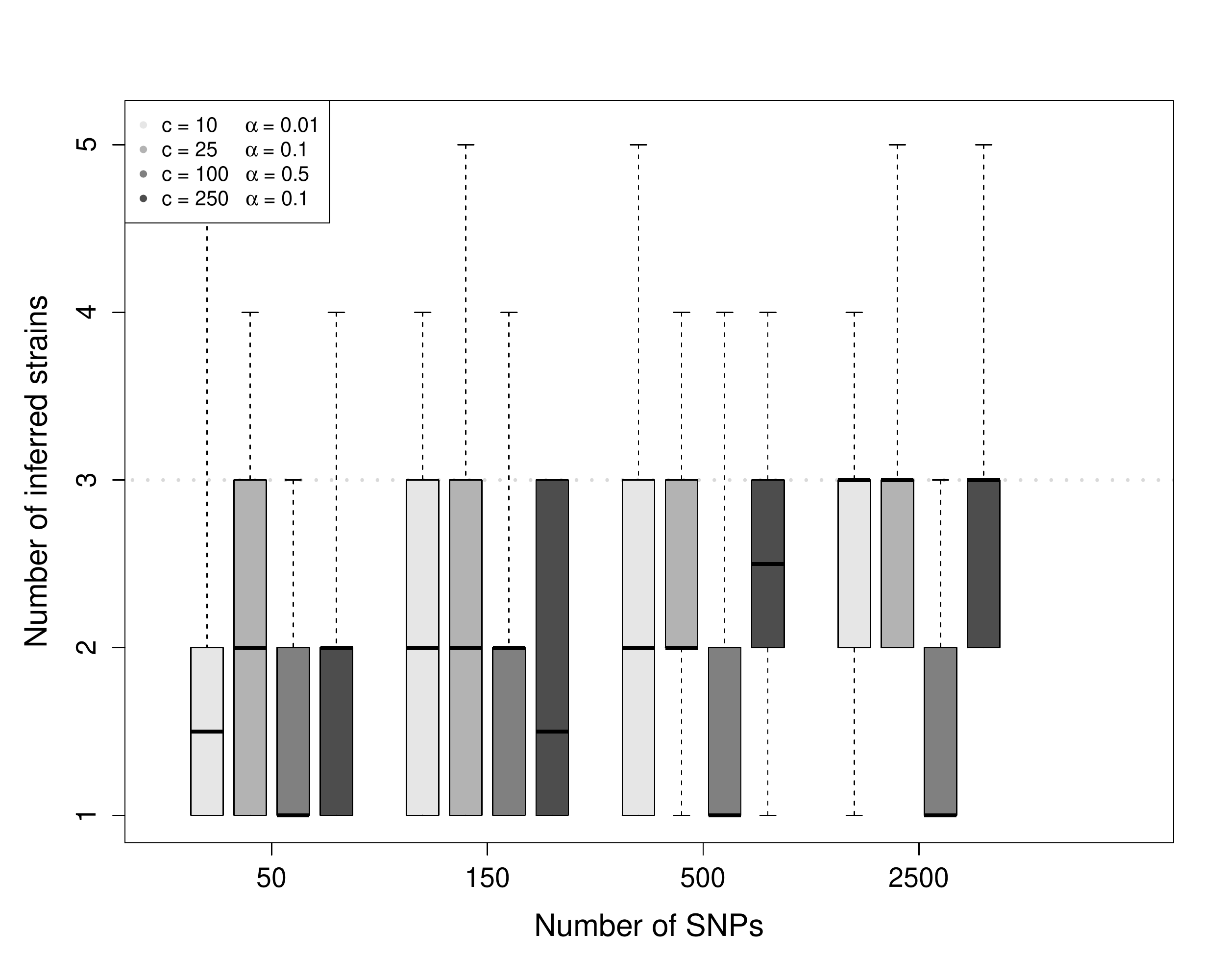} \\
\end{tabular}
\end{center}
\caption{Performance for inference of number of components}
\label{fig:comp_sim}
\end{figure}

Figure \ref{fig:comp_sim} shows performance of the algorithm for inferring the number of components increases in precision with the number of SNPs and the number of reads. Conditional upon $\alpha$, the simulations indicate that the number of SNPs, $M$, to be the largest determinant of performance, with the sum of the read counts, $C$, playing an important supporting role. Inference of the number of underlying strains, $K$, is generally strong for low panmixture levels (small $\alpha$ values), but is noticeably more conservative for $\alpha=0.5$, likely due to the bands becoming increasingly tightly packed as panmixia increases. In general, inference is slightly conservative, likely owing to the BIC estimator's bias toward parsimony \cite{Findley1991}.
\subsubsection*{Parameters}
\begin{figure}[!ht]
\centering
\begin{subfigure}[b]{0.5\textwidth}
                \includegraphics[scale=0.23]{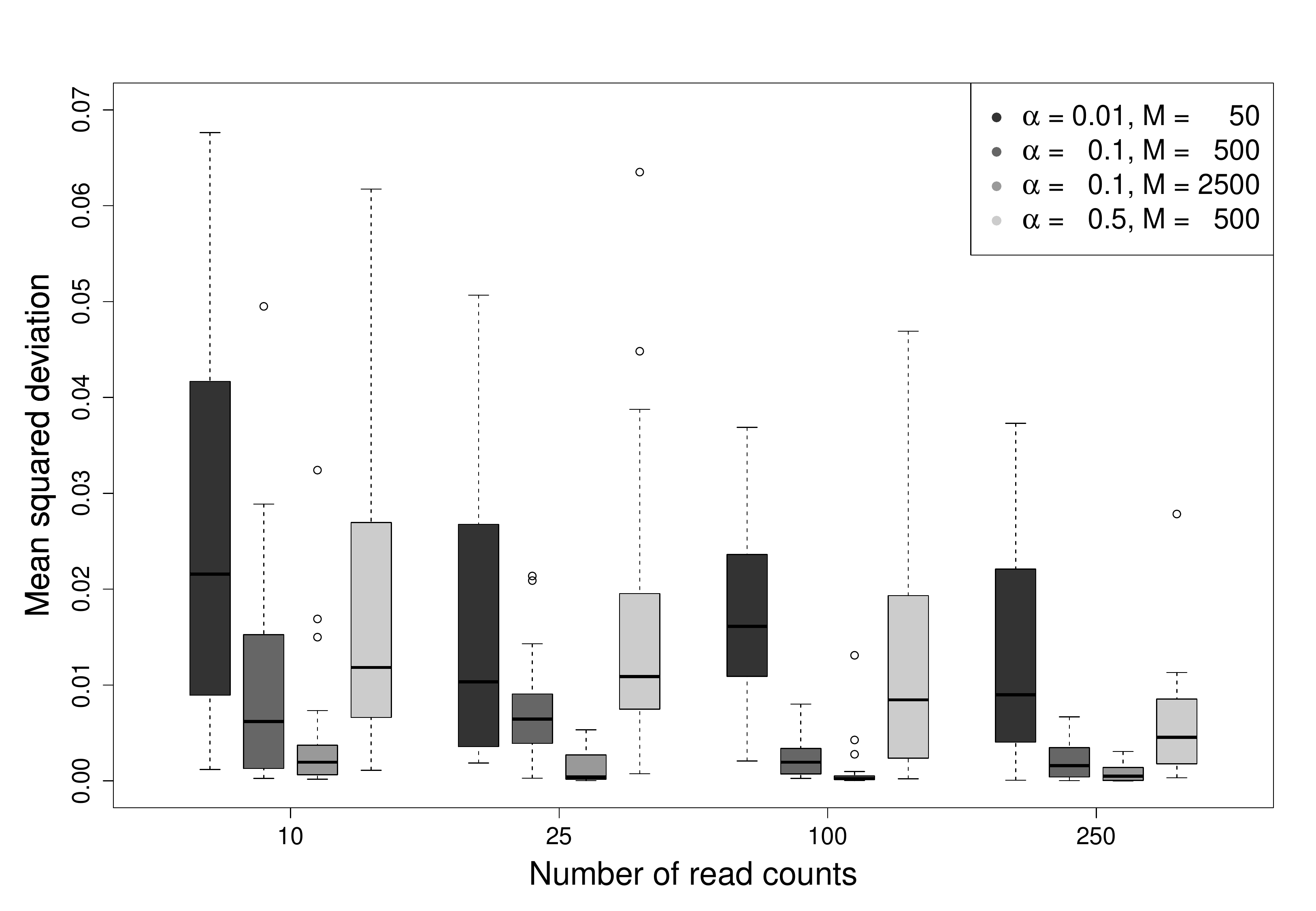}
	           \caption{}
                 \label{fig:gull}
        \end{subfigure}%
\begin{subfigure}[b]{0.5\textwidth}
                \includegraphics[scale=0.23]{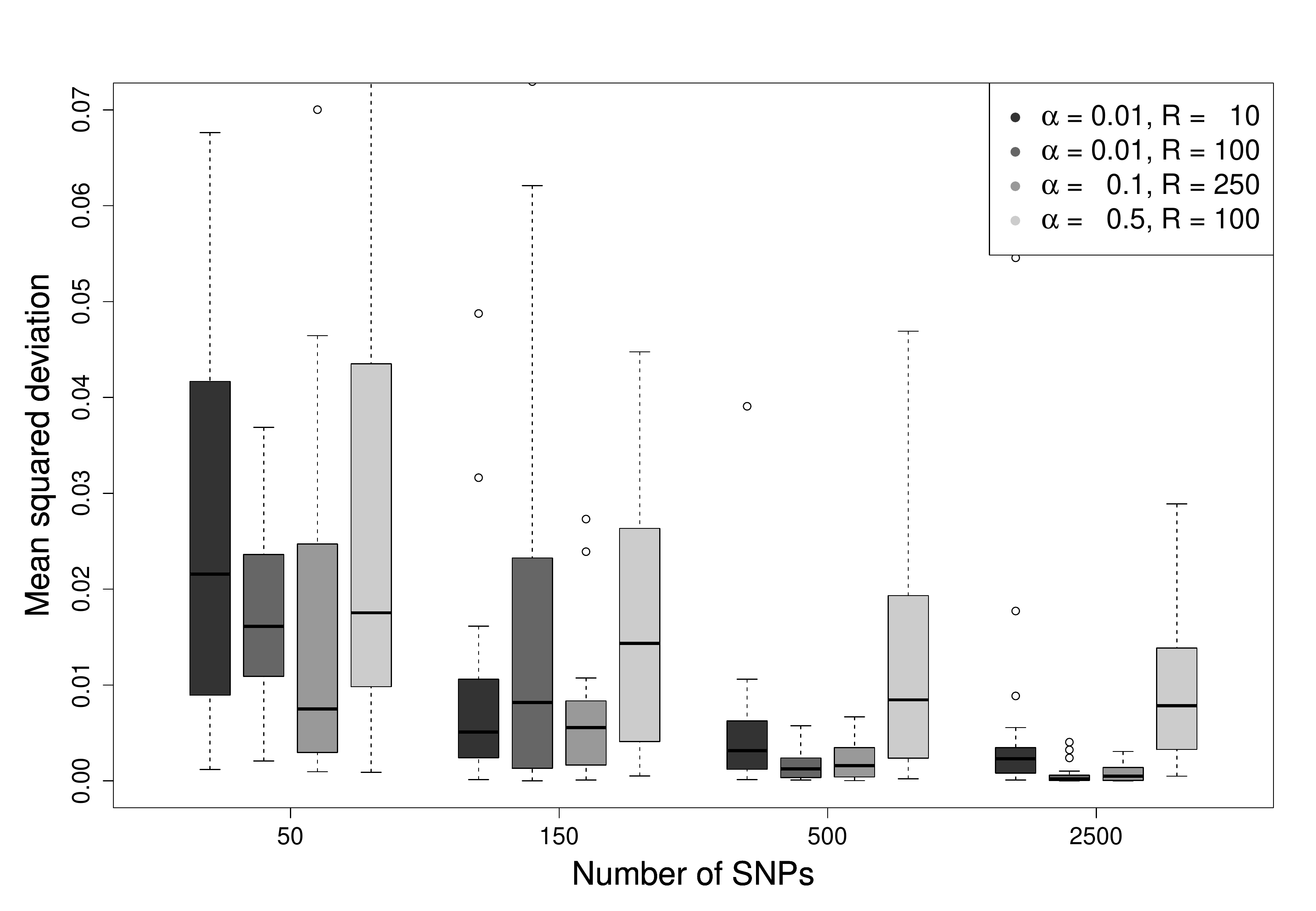}
	           \caption{}
                                \label{fig:gull}
        \end{subfigure} \\
\begin{subfigure}[b]{0.5\textwidth}
                \includegraphics[scale=0.23]{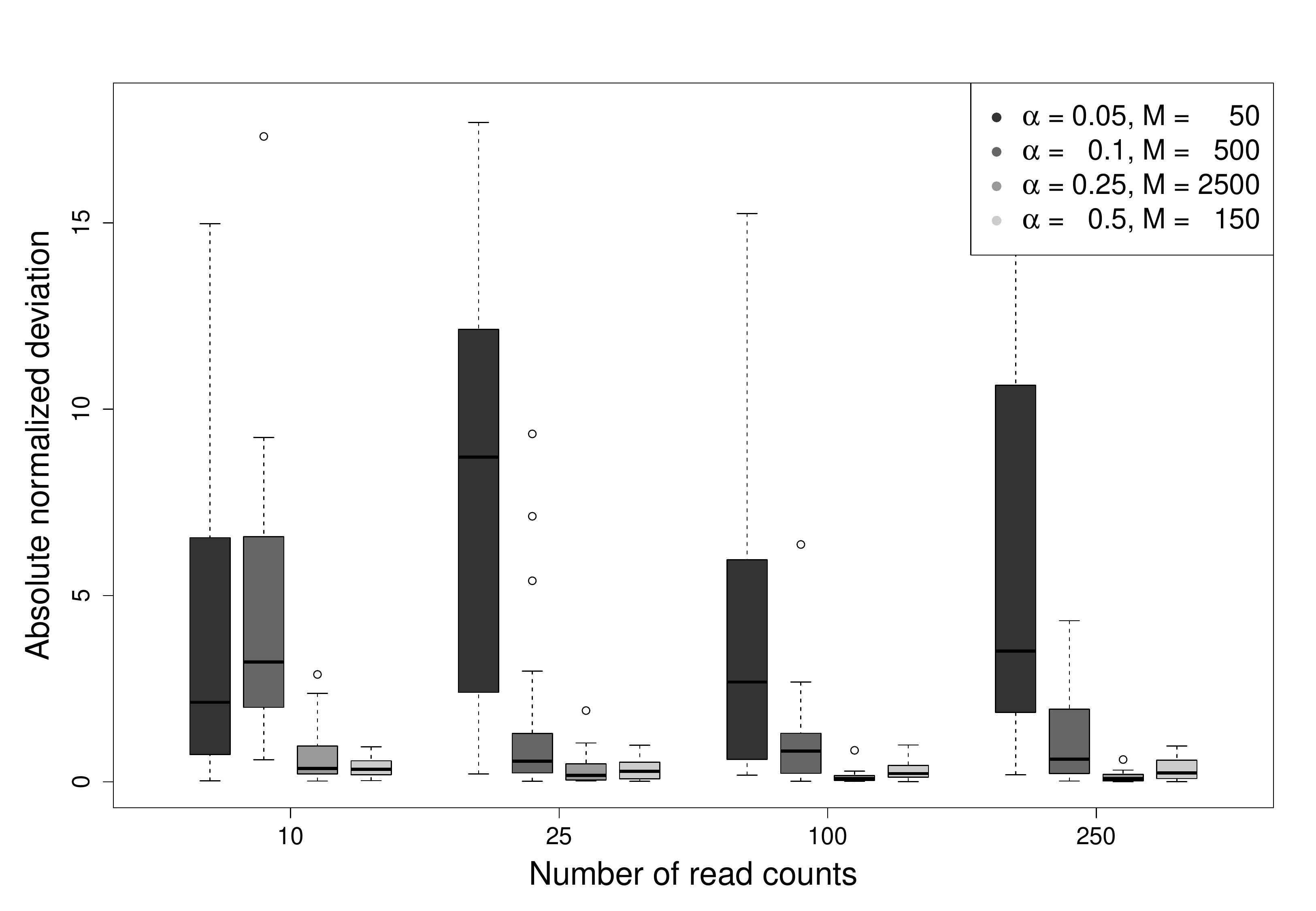}
             	           \caption{}
             	              \label{fig:gull}
        \end{subfigure}%
\begin{subfigure}[b]{0.5\textwidth}
                \includegraphics[scale=0.23]{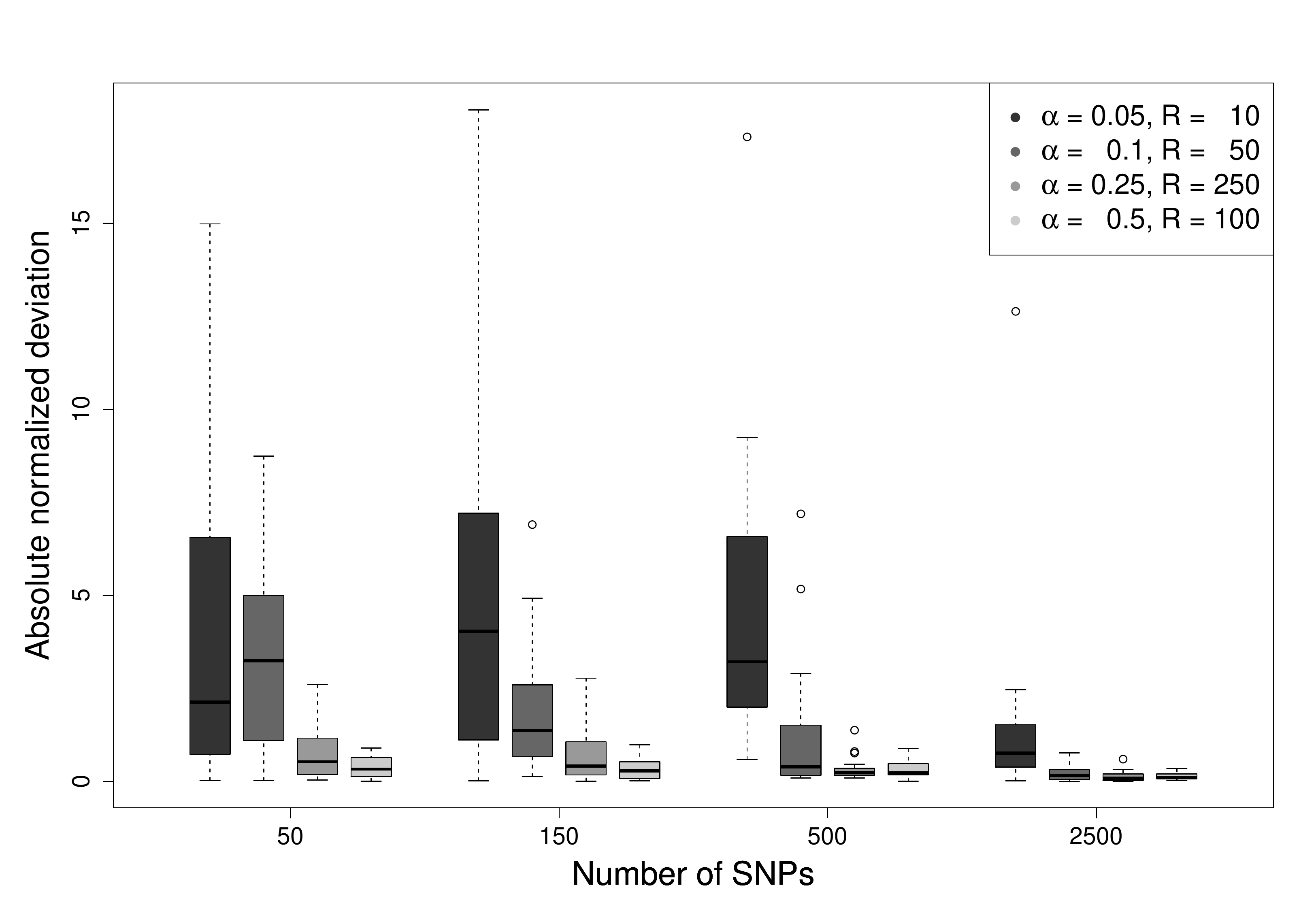}
	           \caption{}
	                           \label{fig:gull}
        \end{subfigure}%
\caption{Performance for parameter inference: (a) mean squared deviation for $\mathcal{W}$ by number of read counts; (b) mean squared deviation for $\mathcal{W}$ by number of SNPs; (c) absolute normalized deviation for $\alpha$ by number of read counts; and (d) absolute normalized deviation for $\alpha$ by number of SNPs.}
\label{fig:comp_sim_para}
\end{figure}

Figure \ref{fig:comp_sim_para} shows similar performance for inference of the strain proportions $\mathcal{W}$ and $\alpha$. For $\mathcal{W}$, we report the mean squared deviation. For $\alpha$, we report the absolute normalized deviation to account for relative difference from the true value. For both parameters, we observe that the number of SNPs is the strongest determinant of accuracy, with $M = 150$ ensuring moderately strong performance. High $\alpha$ moderately decreases the quality of inference for the strain proportions.

\subsection*{Clinical samples from northern Ghana}
\begin{figure}[!ht]
\begin{center}
\begin{tabular}{cc}
\includegraphics[scale=0.37]{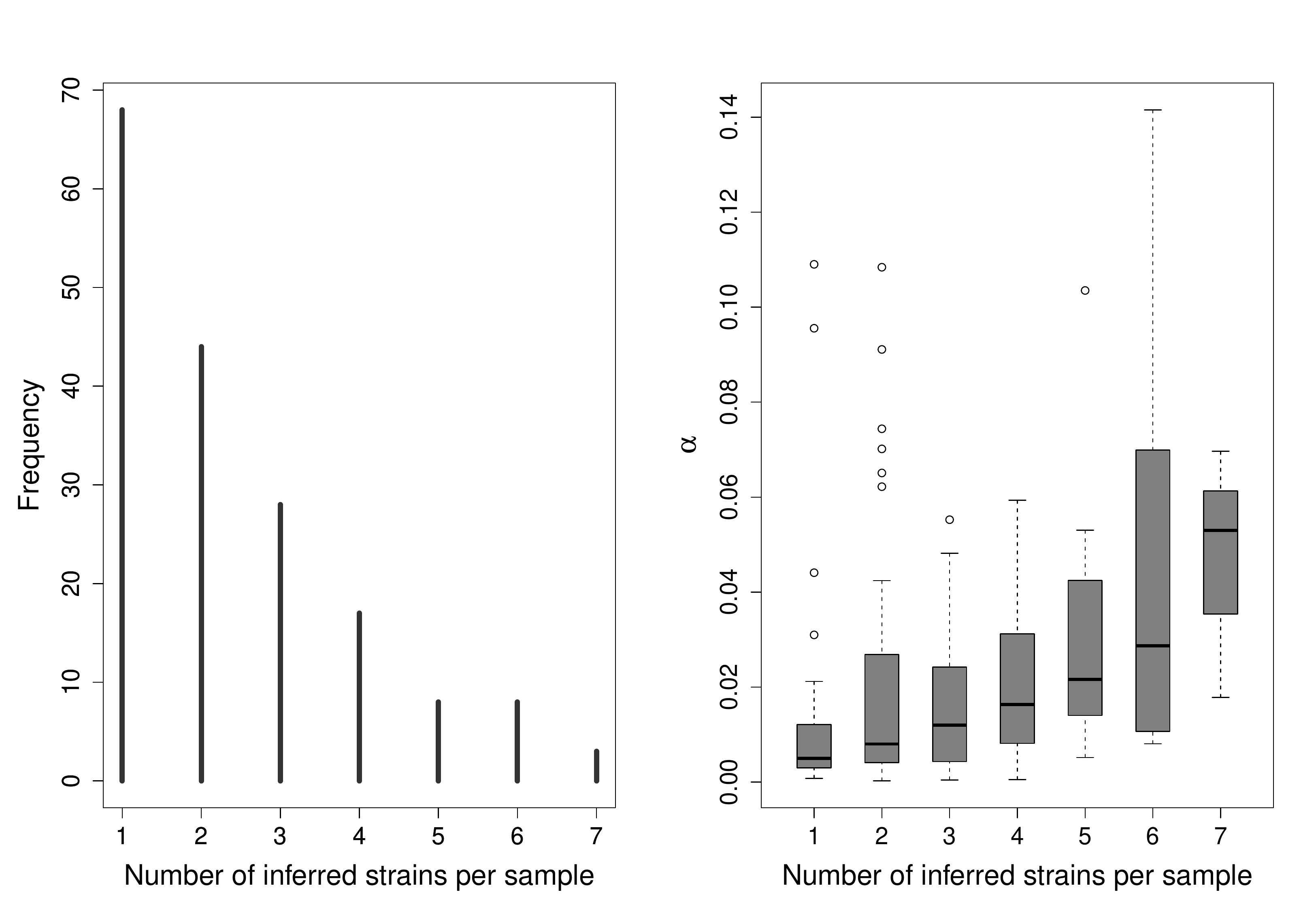} &
\end{tabular}
\end{center}
\caption{The frequency of number of inferred strains per sample (lef) and the posterior median value of $\alpha$ by the number of inferred strains (right).}
\label{fig:portrait}
\end{figure}

Applying the algorithm to the $168$ high-quality samples from KND, we observe $K$ range $1$ to $7$, with $\alpha$ falling between $0$ and $0.14$, and a moderate correlation between $K$ and $\alpha$ (Figure \ref{fig:portrait}). The largest subset of samples were unmixed, with $K=1$ and $\alpha < 0.01$, though the majority of samples exhibit moderate levels of mixture, with $K=2,3,4$ and $0.01\le \alpha \le 0.03$. A small number of samples exhibit complex mixtures, with $K>4$ and $\alpha$ typically greater than $0.02$. These results confirm the presence of mixtures within Pf clinical isolates, but also indicate, possibly more complex patterns involving interactions between the number of dominant strains and the degree of panmixia.

We observe that for most samples the $95\%$ credible interval for $\alpha$ is within a small percentage of the median value. For $\mathcal{W}_i$, we observe a similarly tight posterior distribution, particularly for samples with $K\le3$. The posterior uncertainty increases together with increasing $K$ and increasing $\alpha$. For a small number of samples, the model initially produced unusually low values for $\nu$, indicating a bimodal Beta-binomial distribution inconsistent with a mixture of strains and consequently suspect values for $K$. For these, we bounded $\nu$ such that it ensured a unimodal distribution and then recovered results consistent  the remaining samples.

\begin{figure}[!ht]
\begin{center}
\begin{tabular}{cc}
\includegraphics[scale=0.37]{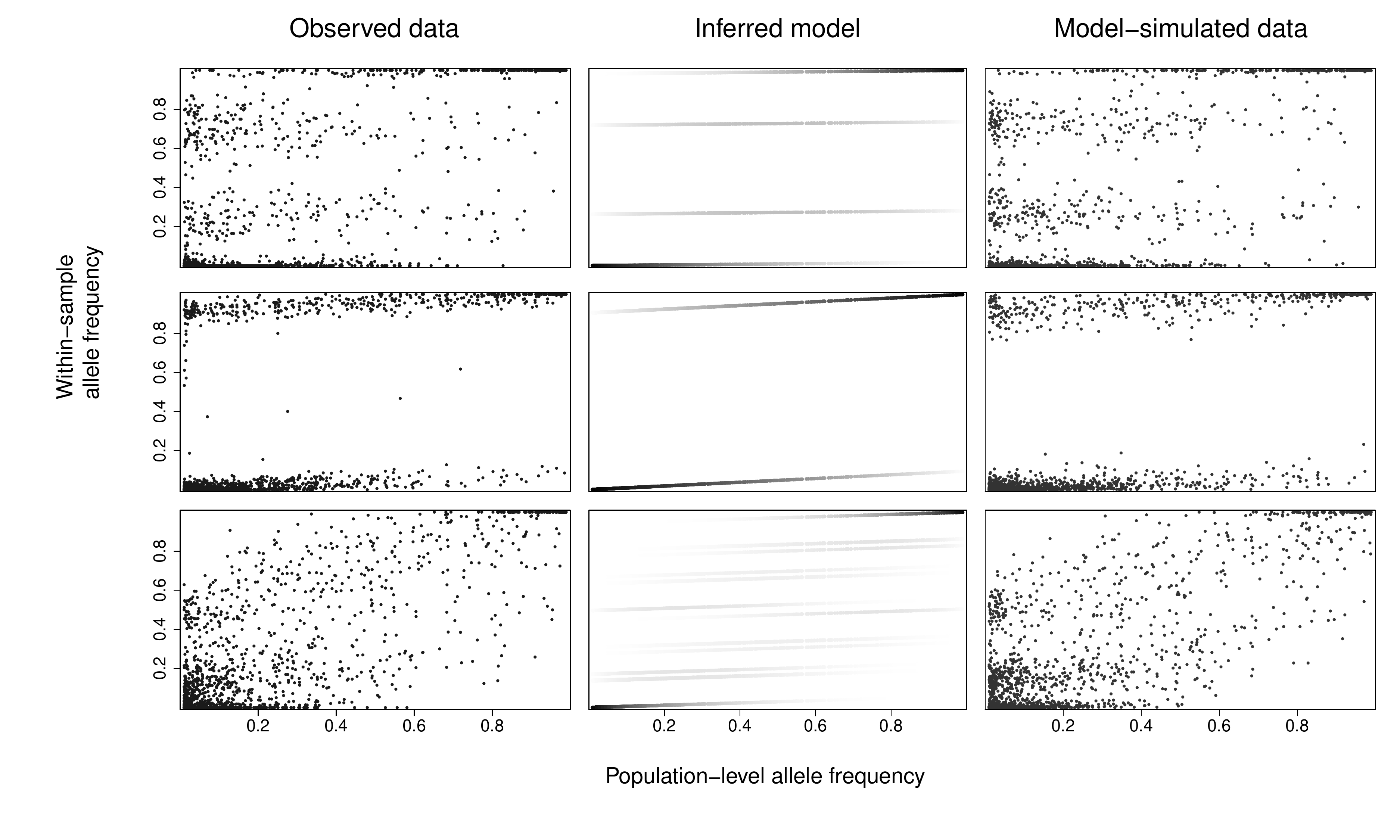} &
\end{tabular}
\end{center}
\caption{Examples of real data. For three samples (rows), we present the observed data WSAF plotted against the PLAF (first column), a diagram of the inferred model indicating the bands, proportions, and $\alpha$ (second column), and data simulated under the inferred model. $\alpha$ and $\mathcal{W}$ are the maximum \emph{a posteriori} values. In the second column, the model's PLAF-varying mixture densities are shown in grey scale, with black equal to one. }
\label{fig:some}
\end{figure}

\begin{figure}[!ht]
\begin{center}
\begin{tabular}{cc}
\includegraphics[scale=0.37]{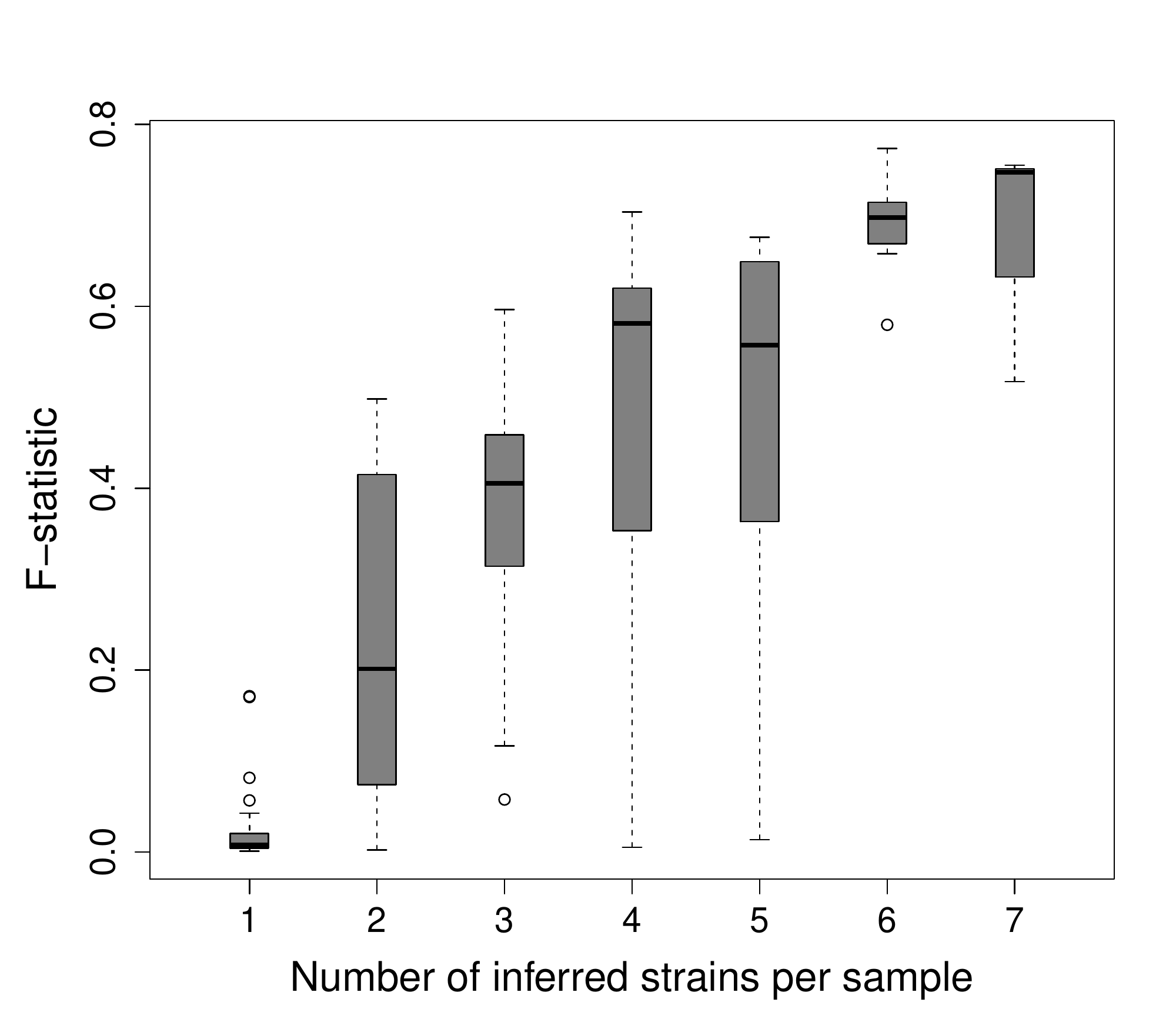} &
\end{tabular}
\end{center}
\caption{Boxplot of the F-statistic inbreeding coefficient ($1-F_{is}$) for each sample grouped by the number of inferred strains. }
\label{fig:f_stat}
\end{figure}

To  visually inspect the quality of the results, we generate figures for each of the samples showing the observed WSAF and PLAF data, the inferred model structure, and data simulated under the inferred model following the observed PLAF. We show examples of these plots for three typical samples in Figures \ref{fig:some}. Nearly all samples (158/168), across all different mixture patterns, show strong visual correspondence between the observed and model-simulated data. We also observe a strong correlation between the inferred number of components and a quasi-maximum likelihood estimate for the inbreeding coefficient for each sample (Figure \ref{fig:f_stat}) \cite{O'Brien2014}.

For each sample, we compare the full model to two reduced versions under the restrictions $\alpha=0$ and $K=1$, respectively. These restrictions correspond to the cases where the model becomes the simple mixture model, with $2^K$ bands but no admixture with the local population, and the panmixture model, where there is a single strain with some local population admixture, respectively. For numerical stability reasons, we set the former restriction as $\alpha=0.001$. For $60\%$ of samples (108/168), the BIC selects the full model over either of the restricted models. For 58 samples the BIC criterion selected the $K=1$ restriction, while for 23 samples it selected $\alpha=0$. Taken in aggregate across all samples, the criterion overwhelmingly selects the full model over either of the restricted models.

 \section*{DISCUSSION}

The model captures two dimensions of within-sample mixture for \emph{P. falciparum} that had previously modeled separately: the number of strains and the degree of admixture with the local population. Evidenced by the comparison of the full model with restricted sub-models, this approaches provides a marked improvement over both more restricted approaches in capturing the structure of mixture in clinical samples. While the model provides a more involved qualitative understanding of the samples, the strong correlation between the inbreeding coefficient and the inferred number of strains shows that the model produces results consistent with previous methods.


In order to perform inference, the model makes a number of simplifying assumptions that may be violated in practice. The model presumes that SNPs are unlinked and consequently independent for the purpose of calculating the likelihood. Given the high recombination rate of \emph{P. falciparum} this assumption may hold for the majority of pairs of SNPs, but neglects correlations that appear locally ($\sim$ 10 kB). However, we expect that this independence assumption serves to moderately weaken the inferential power of the model rather than cause any type of bias since it fails to include possibly informative data, rather than posit a possibly misspecified model. More problematic is the model's implicit assumption of limited population structure. In the case of the KND samples, and perhaps in much of west Africa, this assumption appears supported \cite{Anderson2000,Manske2012}. In other contexts, specifically south-east Asia, recent population bottlenecks and selection suggest that this assumption will be violated \cite{Miotto2013}. The consequences on this model inference are unknown but can likely be partially resolved with appropriate simulation studies.  

The model presents an important new tool for interrogating the biology of clinical Pf infections. In particular, how the number of component strains and the panmixia coefficient relate to the infection parameters, such as seasonality, transmission intensity, and outcrossing, and evolutionary parameters such as the rate of change within sections of the Pf genome, could have implications for understanding the genetic epidemiology of Pf. The model also presents a means for clarifying the poorly detailed structure of intra-host infection dyanmics, such as strain selection or density-dependent selection \cite{Kwiatkowski1991}, by resolving how the number of strains, the mixture proportions, and the panmixia coefficient change within the course of an infection or in response to drug intervention.

An unexpected structural consequence of the model is that power to infer additional strains diminishes as the panmixia coefficient ($\alpha$) increases. This results from the simplifying assumption that $1-\alpha$ of the reads come from the dominant strains while the remaining $\alpha$ fraction are sampled randomly from the local population. Geometrically, we see that as $\alpha$ increases that the bands will get progressively closer together as they approach panmixia, making them harder for the model to distinguish. Also, as $\alpha$ increases to one, the fraction of reads representing the dominant strains diminishes, reducing power to infer these components. We observe this pattern strongly in simulations (Figure \ref{fig:comp_sim_para}): for $\alpha = 0.5$ or greater, the model consistently infers too few components. While this deficiency may be overcome in some fashion with additional SNPs or read counts, the geometry indicates there may be fundamental limits on any model's ability to discriminate the true number of components in the high panmixia regime. 

In principle, the model can be explicitly tested against experiment. Laboratory facilities with the capacity to store many field strains ($> 100$) could generate artificial samples in an experimental analog of our simulation procedure, as follows. Starting with $N$ unmixed strains, identified using inbreeding coefficients, they could create mixtures, they would need to first fix the required sequencing volume as $\eta$, and the parameters for panmixia ($\alpha$), number of component strains ($K$), and their mixture parameters, $\mathcal{S}$ and $\mathcal{W}$. For the finite mixture component, they would then combine volumes of $\eta \cdot \mathcal{W}$ from the $K$ strains. For the panmixture component, they would then fix some large number but experimentally feasible number of strains (say $100$) and randomly sample from all of them a volume of $\eta/100$. Finally, combining these into final sample and applying WGS sequencing, will yield data that we hypothesize will closely follow the integrated model outlined above, with $\nu$ capturing the experimental variation. Naturally, consistent results would indicate the sufficiency of the model, but not it's necessity, holding out the possibility of a more minimal description. These results could be further compared against other next-generation technologies, such as single-cell sequencing, that have been deployed to understand Pf clinical mixtures \cite{Nair2014}. 

The method works efficiently in practice (the properties of a single sample can be inferred in minutes on a standard laptop) but a number of possible improvements could strengthen its statistical performance. Most immediately, creating a full Bayesian approach rather than the parallelizing implementation here - while likely not improving the parametric inference for individual samples - would provide the full posterior distribution across all samples for more considered model comparison. In that same line, more refined approaches to inferring the number of strains within samples, either via a reversible jump MCMC approach or methods for rigorously estimating Bayes factors \cite{Green1995}, would provide researchers more power to resolve the structure of mixture, though likely at the cost of significantly more computation.  

The model does not perform haplotype phasing to resolve the sequence of the underlying strains \cite{Stephens2001,Howie2012,O'Brien2014b}. The analysis here suggests that a method for estimating haplotypes would be straight-forward for some samples (unmixed ones, for instance) but difficult if not impossible for others (when $\alpha$ is greater than $0.5$). Researchers may be particularly interested in whether, in these phased samples, particular stretches of the genome appear more or less frequently in the dominant strains than others, indicating immunological or environmental selection. This is also an avenue for statistical development.

The two phenomenologies of mixture that the model captures - a finite mixture of distinct strains and an inbred population admixture - cannot be immediately associated with any specific aspect of the infection process. A number of variables appear plausible in determining these relationships, including transmission intensity, the length of infection, the immunological status of the infected individual, and within-host density dependent selection. Together with WGS data, this new approach can serve as a means for biological researchers to directly resolve these hypotheses and resolve the consequence of mixture in \emph{P. falciparum} infections.  

\subsection*{Authorship}
JO designed and implemented the study and wrote the manuscript. ZI assisted in the design of the study and commented on the manuscript. LA-E collected the data and commented on the manuscript and the study.

\section*{Tables}

\begin{table}[!ht]
\begin{center}
\begin{tabular}{l|l}
Parameter & Definition \\
\hline $N$ & Number of samples \\
$M$ & Number of SNPs \\
$K$ & Number of strains \\
$i = 1,\cdots,N$ & Index for samples \\
$j = 1,\cdots,M$ & Index for SNPs \\
$r = 1,\cdots 2^K$ & Index for bands / strain mixtures \\
$p_j$ & (Non-reference) allele frequency for SNP $j$ \\
$\mathcal{P} = [p_j]$ & The PLAF for all SNPs \\
$\mathcal{Q} = [q_{ij}] $ & Within-sample allele frequency for SNP $j$ in sample $i$ \\
$\alpha$ & Degree of panmixia within a sample, panmixia coefficient\\
$\mathcal{S} = [s_1,\cdots,s_K]$ & Strains in a sample \\
$\mathcal{W} = [w_1,\cdots,w_K]$ & Strain proportions in a sample \\
$\lambda_r$ & Band proportions within sample \\
$\nu$ & Variation parameter for Beta-binomial \\
WSAF & Within-sample allele frequency \\
PLAF & Population-level allele frequency
\end{tabular}
\end{center}
\caption{Parameters and definitions for the model and its description.}
\label{notation}
\end{table}

\begin{table}[!ht]
\begin{center}
  \begin{tabular}{c|rrrr}
Parameter & Values: & & & \\
  \hline   M & 50 & 150 & 500 & 2500\\
  C & 10 & 25 & 100 & 250 \\
  $\alpha$ & 0.01 & 0.1 & 0.5 & \\
  $K$ & 1 & 3 & & \\
    \end{tabular}  
    \caption{Table of simulated parameter values: $C$ the number of read counts while $M$, $K$ and $\alpha$ are as in Table \ref{notation}.}
    \label{table:parameters}
    \end{center}
\end{table}

\newpage

\section*{Figures}

\clearpage
\bibliography{pfmixture_jobrien_arxiv}

\end{document}